\documentclass[twocolumn,prb,superscriptaddress,showpacs]{revtex4-1}
\usepackage{amsmath,amssymb}
\usepackage{graphicx}
\usepackage{verbatim}
\usepackage{amsfonts}
\usepackage{dcolumn}
\usepackage{bm}
\usepackage{color}
\usepackage[colorlinks,citecolor=blue]{hyperref}

\begin{document}

\newcommand{\odiff}[2]{\frac{\di #1}{\di #2}}
\newcommand{\pdiff}[2]{\frac{\partial #1}{\partial #2}}
\newcommand{\di}{\mathrm{d}}
\newcommand{\ii}{\mathrm{i}}
\renewcommand{\vec}[1]{{\mathbf #1}}
\newcommand{\vx}{{\bm x}}
\newcommand{\ket}[1]{|#1\rangle}
\newcommand{\bra}[1]{\langle#1|}
\newcommand{\pd}[2]{\langle#1|#2\rangle}
\newcommand{\tpd}[3]{\langle#1|#2|#3\rangle}
\renewcommand{\vr}{{\vec{r}}}
\newcommand{\vk}{{\vec{k}}}
\renewcommand{\ol}[1]{\overline{#1}}
\newtheorem{theorem}{Theorem}
\newcommand{\comments}[1]{}
\newcommand{\scrap}[1]{{\color{grey}{\sout{#1}}}}

\newcommand{\nts}[1]{[\emph{\color{red}{#1}}]}

\newcommand{\tj}[6]{ \begin{pmatrix}
  #1 & #2 & #3 \\
  #4 & #5 & #6
\end{pmatrix}}

\newcommand {\C}{\textcolor {red}}
\newcommand {\B}{\textcolor {blue}}
\newcommand {\Y}{\textcolor {yellow}}

\title{Topology and criticality in the resonating Affleck-Kennedy-Lieb-Tasaki loop spin liquid states}

\author{Wei Li}
\affiliation{Physics Department, Arnold Sommerfeld Center for Theoretical Physics,
and Center for NanoScience, Ludwig-Maximilians-Universit\"at, 80333 Munich, Germany}
\author{Shuo Yang}
\affiliation{Max-Planck-Institut f\"ur Quantenoptik, Hans-Kopfermann-Str. 1, D-85748 Garching, Germany}
\author{Meng Cheng}
\affiliation{Station Q, Microsoft Research, Santa Barbara, CA 93106, USA}
\author{Zheng-Xin Liu}
\affiliation{Institute for Advanced Study, Tsinghua University, Beijing 100084, China}
\author{Hong-Hao Tu}
\affiliation{Max-Planck-Institut f\"ur Quantenoptik, Hans-Kopfermann-Str. 1, D-85748 Garching, Germany}

\date{\today}
\begin{abstract}
We exploit a natural Projected Entangled-Pair State (PEPS) representation for the resonating Affleck-Kennedy-Lieb-Tasaki loop (RAL) state. By taking advantage of PEPS-based analytical and numerical methods, we characterize the RAL states on various two-dimensional lattices. On square and honeycomb lattices, these states are critical since the dimer-dimer correlations decay as a power law. On kagome lattice, the RAL state has exponentially decaying correlation functions, supporting the scenario of a gapped spin liquid. We provide further evidence that the RAL state on the kagome lattice is a $\mathbb{Z}_2$ spin liquid, by identifying the four topological sectors and computing the topological entropy. Furthermore, we construct a one-parameter family of PEPS states interpolating between the RAL state and a short-range Resonating Valence Bond state and find a critical point, consistent with the fact that the two states belong to two different phases. We also perform a variational study of the spin-1 kagome Heisenberg model using this one-parameter PEPS.
\end{abstract}
\maketitle

\section{Introduction}
The quest for physical systems harboring quantum spin liquid states has been a long-standing challenge in condensed matter physics~\cite{Balents-2010}. These states do not exhibit any symmetry breaking and thus admit no descriptions in terms of local order parameters. In contrast to conventional ordered states, spin liquids may have topological order~\cite{Wen-1990, Wen-Niu-1990} and support exotic fractional excitations. Among various theoretical constructions, a class of gapped $\mathbb{Z}_2$ spin liquids has been well-studied and recently receives strong numerical support~\cite{White-2011,Schollwoeck-2012, Jiang-2012a, Wang-2011} as candidate ground states of some physical spin Hamiltonians on frustrated lattices.
Their ground-state wave functions contain long-range entanglement and the low-energy effective theory is conveniently described by an emergent $\mathbb{Z}_2$ gauge theory~\cite{Kitaev-2003,Senthil-2000}, which also offers an intuitive picture for the highly-entangled ground state: a soup of fluctuating $\mathbb{Z}_2$ electric field lines \cite{Levin-2005}.
Moreover, if the spin system has $\mathbb{SO}(3)$ symmetry, the $\mathbb{Z}_2$ gauge charges and fluxes in different spin liquid states may transform as linear (integer spin) or projective (half-integer) representations under the action of the symmetry group, i.e., they may belong to different classes of Symmetry Enriched Topological (SET) phases~\cite{Essin-2013, Mesaros-2013, Hung-2013, Lu-2013,Fidkowski-unpub}.

\begin{figure}[htpb]
  \begin{center}
	\includegraphics[width=0.9\columnwidth]{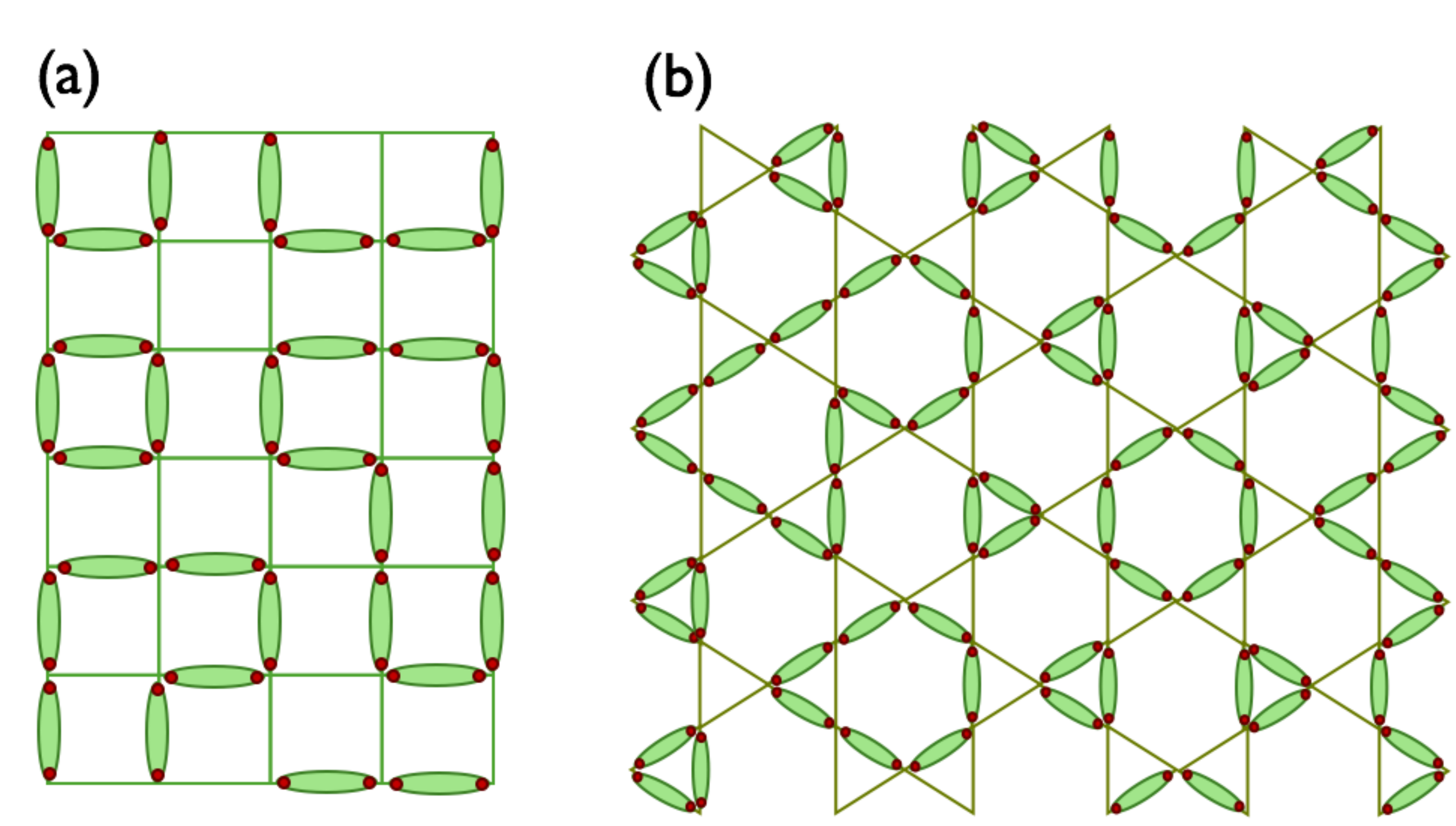}
  \end{center}
  \caption{(Color online) Illustration of typical AKLT-loop configurations in the RAL states on (a) square and (b) kagome lattices. Each green long oval represents a valence bond inside, while the small red dots represent the constituting spin-1/2's. }
  \label{fig:RLS}
\end{figure}

In this paper, we characterize a type of spin-$1$ spin liquids~\cite{Liu-2010, FWang-2011, Serbyn-2011,Cenke-2012,Bieri-2012,ZXLiu-2012,Tu-2013,ZCai-2007}. The representative wave function of these states is a superposition of strongly fluctuating, fully packed loops (see Fig.~\ref{fig:RLS}), where each loop carries an Affleck-Kennedy-Lieb-Tasaki (AKLT) state~\cite{Affleck-1987}. We call this state a resonating AKLT-loop (RAL) state following Ref. [\onlinecite{HYao-2010}], where it was considered as an example of nontrivial SET phases. We provide an explicit Projected Entangled-Paired State (PEPS) representation of the state~\cite{Verstraete-2004, Verstraete-2006}, which allows us to exploit efficient PEPS-based analytical and numerical techniques to characterize these wave functions on various lattices. We demonstrate that on bipartite lattices (e.g. square and honeycomb lattices) these states have algebraically decaying dimer-dimer correlations and exponentially decaying spin-spin and quadrupole-quadrupole correlations, indicating a gapless spin liquid. On the non-bipartite kagome lattice, all correlation functions we have tested (spin, dimer and quadrupole) decay exponentially, whose correlation lengths do not increase with the system size. Additionally, we compute the topological entanglement entropy~\cite{Kitaev-2006, Levin-2006} of these wave functions on a long cylinder and obtain a universal value $\gamma \approx \ln 2$. All these results are indicative of a gapped $\mathbb{Z}_2$ spin liquid.

The AKLT state is a paradigmatic example of symmetry-protected topological phase in one dimension~\cite{Gu-2009,Pollmann-2010,Schuch-2011}, featuring degenerate spin-$1/2$ excitations on boundaries protected by the $\mathbb{SO}(3)$ spin rotation symmetry. With no surprise, the RAL state also exhibits interesting symmetry properties. Namely, terminations of the loops always carry  spin-$1/2$ degrees of freedom. In contrast, the spin-1 short-range RVB state, unlike its spin-1/2 counterpart\cite{Anderson-1973, Rokhsar-1988}, does not support deconfined spin-$1/2$ excitations. We discuss the manifestation of this interesting symmetry enrichment in the construction of the degenerate topological sectors on an infinitely long cylinder. As a demonstration, we construct a one-parameter family of PEPS states interpolating between the RAL state and the RVB state. We find a critical point along the path, where the dimer-dimer correlation function decays algebraically. We also use these states as variational wave functions for the \mbox{spin-1} kagome Heisenberg model and find an upper bound for the ground-state energy.

The rest of the paper is organized as follows. In Sec. II, we introduce the PEPS representations of the RAL states and related PEPS algorithms for extracting physical quantities. In Sec. III, we show that the RAL states are critical on square and honeycomb lattices. In Sec. IV, we exploit a combination of analytical and numerical techniques to provide evidence that the RAL state on a kagome lattice is a gapped spin liquid with $\mathbb{Z}_2$ topological order. In Sec. V, a one-parameter PEPS based on the RAL state are utilized for a variational study of the spin-1 Heisenberg model. Lastly, the Sec. VI is devoted to a summary.

\section{PEPS representation and algorithms}
In this section, we introduce the PEPS representation of the RAL state and describe the PEPS algorithms utilized to extract physical quantities.

\subsection{PEPS representation for RAL states}
\begin{figure}[htpb]
  \begin{center}
	\includegraphics[width=0.9\columnwidth]{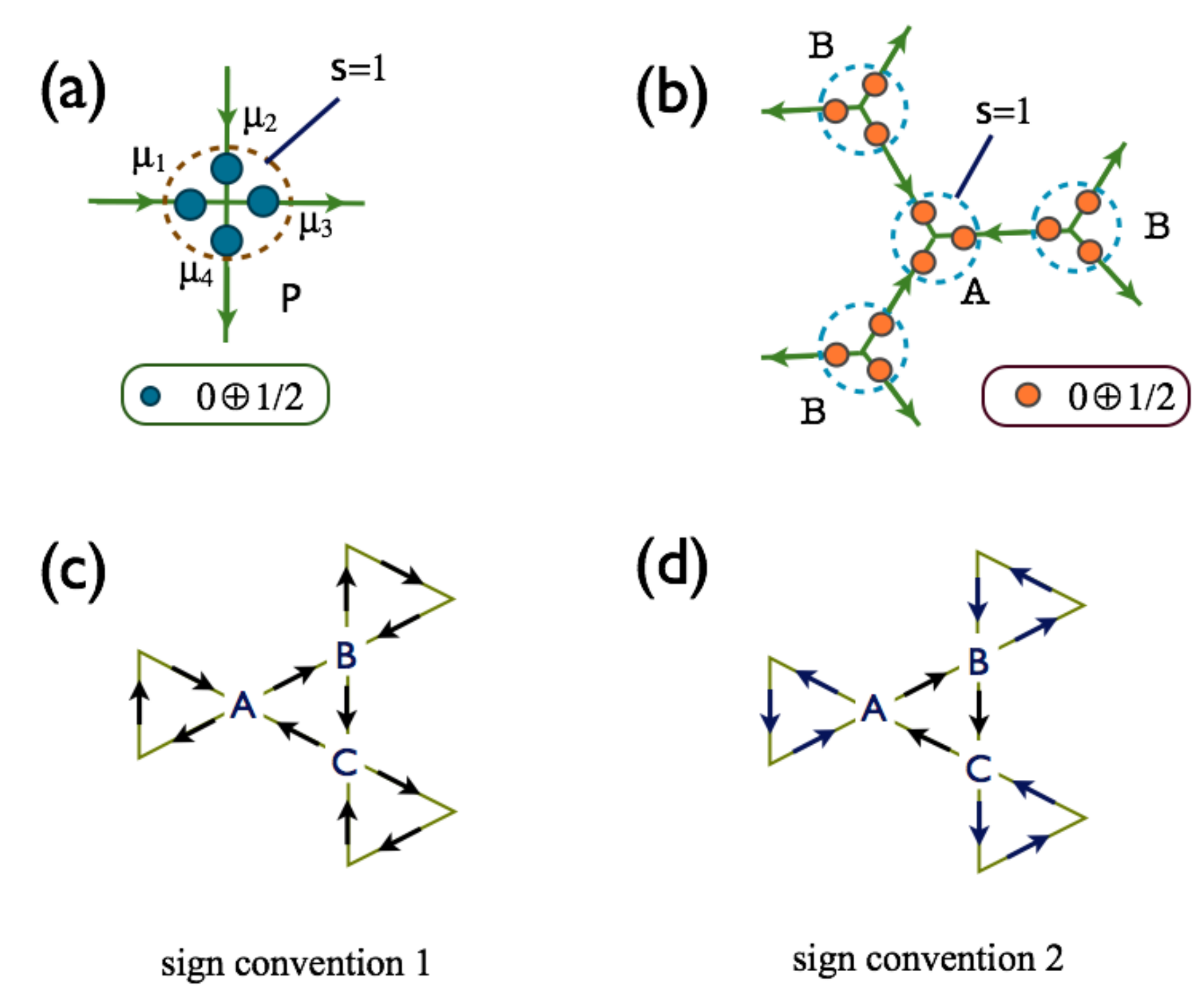}
  \end{center}
  \caption{(Color online) (a) The PEPS projector $P$ for the RAL state on square lattice, mapping four virtual particles ($0 \oplus\frac{1}{2})^{\otimes4}$ onto the physical spin-1 Hilbert space. (b) PEPS representation of the RAL state on a honeycomb lattice. (c,d) Two sign conventions for the RAL states on a kagome lattice. Arrows denote the orientations of the virtual singlets.}
  \label{fig:peps}
\end{figure}

Let us introduce the RAL state on a square lattice. In this case, we associate each lattice site with four virtual particles, each of which has three basis states, $\ket{0}\equiv\ket{\emptyset}$, $\ket{1}\equiv\ket{\uparrow}$, and $\ket{2}\equiv\ket{\downarrow}$, representing a spin-0 vacancy and a spin-$1/2$ doublet, respectively. In the PEPS language that we shall extensively use below, the virtual Hilbert space is denoted by the $\mathbb{SU}(2)$ representation $0 \oplus  1/2$ (i.e. direct sum of spin 0 and spin 1/2), whose dimension is called bond dimension, denoted by $D$. Thus, we have $D=3$ in the present case. The RAL state can be constructed as
\begin{equation}
| \Psi_{\rm{RAL}} \rangle = \bigotimes_{i=1}^N P_i \, \bigotimes_{\langle ij \rangle} |\epsilon \rangle_{ij},
\label{eq:PEPS}
\end{equation}
where $|\epsilon \rangle_{ij}$ is a maximally entangled bond state connecting the virtual particles between neighboring sites $i$ and $j$, defined by $|\epsilon \rangle= |0,0\rangle + |1,2\rangle - |2,1\rangle$, and $P_i$ is a local projector acting on site $i$, which maps two virtual spin-1/2 states, out of the four virtual particles, onto the physical spin-1 Hilbert space [see Fig.~\ref{fig:peps}(a) for an illustration]. More explicitly, $P$ can be written as
\begin{equation}
P = \sum_{1\le l < l' \le 4} P_{l,l'},
\label{eq:PEPSprojector}
\end{equation}
where, for instance, $P_{1,2}$ is given by
\begin{equation}
P_{1,2} = \sum_{m} \sum_{\mu_1 \mu_2 \mu_3 \mu_4} C_{\mu_1, \mu_2}^m \delta_{\mu_3, 0} \delta_{\mu_4,0} |m\rangle \langle \mu_1,\mu_2,\mu_3,\mu_4|.
\label{eq:projector}
\end{equation}
Here $m \in \{\pm1,0\}$ and $\mu_1,\mu_2,\mu_3,\mu_4 \in \{0,1,2\}$ denote the physical and virtual Hilbert spaces, respectively. $C_{\mu_1, \mu_2}^m$ is the Clebsch-Gordan (CG) coefficient symmetrizing two spin-1/2 particles into a physical spin-1, with nonvanishing coefficients being $C_{1, 1}^1 = C_{2, 2}^{-1} = 1$ and $C_{1, 2}^0 = C_{2, 1}^{0} = 1/\sqrt{2}$.

To uniquely define the wave function, one needs to specify the orientation of the virtual singlets $|\epsilon \rangle$ up to a gauge choice. For square and honeycomb lattices, we use the sign conventions shown in Fig.~\ref{fig:peps}(a) and (b), where the singlets are oriented according to the arrows. For the kagome lattice, we consider two inequivalent sign conventions shown in Fig. \ref{fig:peps}(c) and (d).

Up to now, we have introduced the RAL state (\ref{eq:PEPS}) in the PEPS language. If we substitute (\ref{eq:PEPSprojector}) into (\ref{eq:PEPS}) and expand the product $\bigotimes_{i=1}^N P_i$, the resonating AKLT-loop picture of (\ref{eq:PEPS}) shown in Fig. \ref{fig:RLS} also becomes transparent, as each configuration contains different patterns of fully packed spin-1 AKLT loops covering the whole lattice. Since the projector in (\ref{eq:PEPSprojector}) gives the \emph{same} weight to all allowed ways of combining two spin-1/2's into a physical spin-1, the wave function (\ref{eq:PEPS}) can be viewed as an \emph{equal} weight superposition (up to a sign depending on the sign convention) of all possible AKLT-loop configurations, which was first proposed in Ref. [\onlinecite{HYao-2010}].

At this point, we note that our PEPS representation naturally generalizes to more complicated resonating loop states, where the loops carry other 1D matrix-product states (e.g. $\mathbb{SO}(2n+1)$ AKLT state \cite{Tu-2008}).

\subsection{RAL states on the kagome lattice}
\label{sec:kagomeRAL}
\begin{figure}[htpb]
  \begin{center}
	\includegraphics[width=0.85\columnwidth]{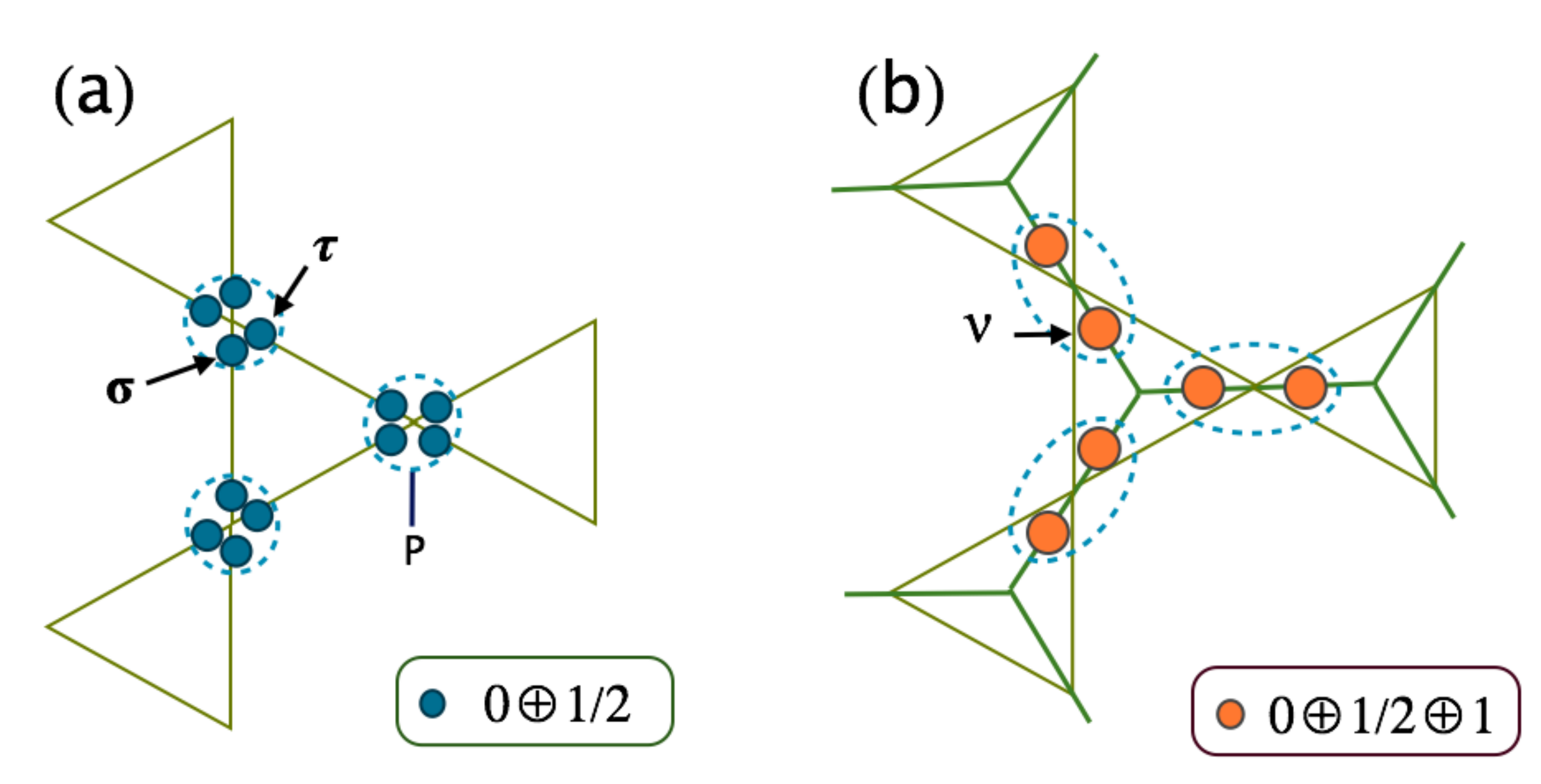}
  \end{center}
  \caption{(Color online) (a) The PEPS construction of the pure kagome RAL state. (b) The simplified PEPS construction based on combining virtual particles.}
  \label{fig:peps-kag}
\end{figure}

A straightforward construction of the RAL state on the kagome lattice proceeds along the description in the previous subsection. As shown in Fig. \ref{fig:peps-kag}(a), there are four virtual particles on each vertex, in the representation $0 \oplus 1/2$, and the on-site projection operator $P$ is exactly the same as Eq. (\ref{eq:projector}).

For the RAL state on the kagome lattice, this PEPS representation has a redundancy which can be revealed by the following procedure: we group the two virtual particles belonging to the same triangle (see e.g. $\sigma, \tau$ in Fig.~\ref{fig:peps-kag}(a)) and block them into a single virtual particle (see $\nu$ in Fig.~\ref{fig:peps-kag}(b)). It turns out that not all virtual degrees of freedom survive after this blocking procedure. After removing this redundancy, we find that the new virtual particle $\nu$ is in the $\mathbb{SU}(2)$ representation $0 \oplus 1/2 \oplus 1$ and has dimension $D=6$ (see Fig.~\ref{fig:peps-kag}(b)), instead of $D=9$ from a naive counting of the dimension of the tensor product $(0 \oplus  1/2)\otimes(0 \oplus  1/2)$. Let us note that this procedure is exact and the removal of redundancy allows us to reduce computational costs in our numerical calculations.

Furthermore, in our framework it is straightforward to introduce a one-parameter family of PEPS, which interpolates between the RAL and the spin-1 RVB states. This is done by extending the virtual representation [$\sigma, \tau$ in Fig. \ref{fig:peps-kag} (a)] from $0 \oplus  1/2$ to $0 \oplus 1/2 \oplus 1$. Thus, the virtual bonds in (\ref{eq:PEPS}) are modified as
$|\epsilon_{\rm{mix}} \rangle = \ket{0,0} + \ket{1,2} - \ket{2,1} + (\ket{3,5} - \ket{4,4} + \ket{5,3})/\sqrt{3}$,
where $|3,4,5\rangle$ denotes the three states in the virtual spin-1 space. Accordingly, the projector $P$ in (\ref{eq:PEPSprojector}) has to be modified as
\begin{equation}
P' = (1-\alpha) P + \alpha W,
\label{eq:interpolation}
\end{equation}
where $W$ is a local projector which maps one virtual spin-1 state, out of the four virtual spin-1 particles, onto the physical spin-1 Hilbert space. Its explicit form is given by
\begin{equation}
W = \sum_{l=1}^{4} W_{l},
\end{equation}
where, for instance, $W_1$ is defined as
\begin{equation}
W_1 = \sum_{m} \sum_{\mu_1 \mu_2 \mu_3 \mu_4} C_{\mu_1, 0}^m  \delta_{\mu_2, 0}  \delta_{\mu_3, 0} \delta_{\mu_4,0} |m\rangle \langle \mu_1,\mu_2,\mu_3,\mu_4|.
\end{equation}
Here $C_{\mu, 0}^m$ store the trivial CG coefficients $C_{3,0}^{1} = C_{4,0}^0=C_{5,0}^{-1}=1$. In this one-parameter PEPS, we recover the RAL state (\ref{eq:PEPS}) for $\alpha=0$ and the RVB state for $\alpha=1$. Hence we have a continuous family of ``mixed'' RAL states parametrized by the weight $\alpha$ controlling the density of the spin-1 valence bonds in the mixed loop-dimer configuration. Like in the pure RAL state, we can combine two virtual particles $\sigma, \tau$ into a single virtual particle $\nu$ (see Fig. \ref{fig:peps-kag}). As a result, the virtual particle $\nu$ is also in the representation $0 \oplus 1/2 \oplus 1$ and thus the resulting PEPS has bond dimension $D=6$.

\begin{figure}[htpb]
  \begin{center}
	\includegraphics[width=0.95\columnwidth]{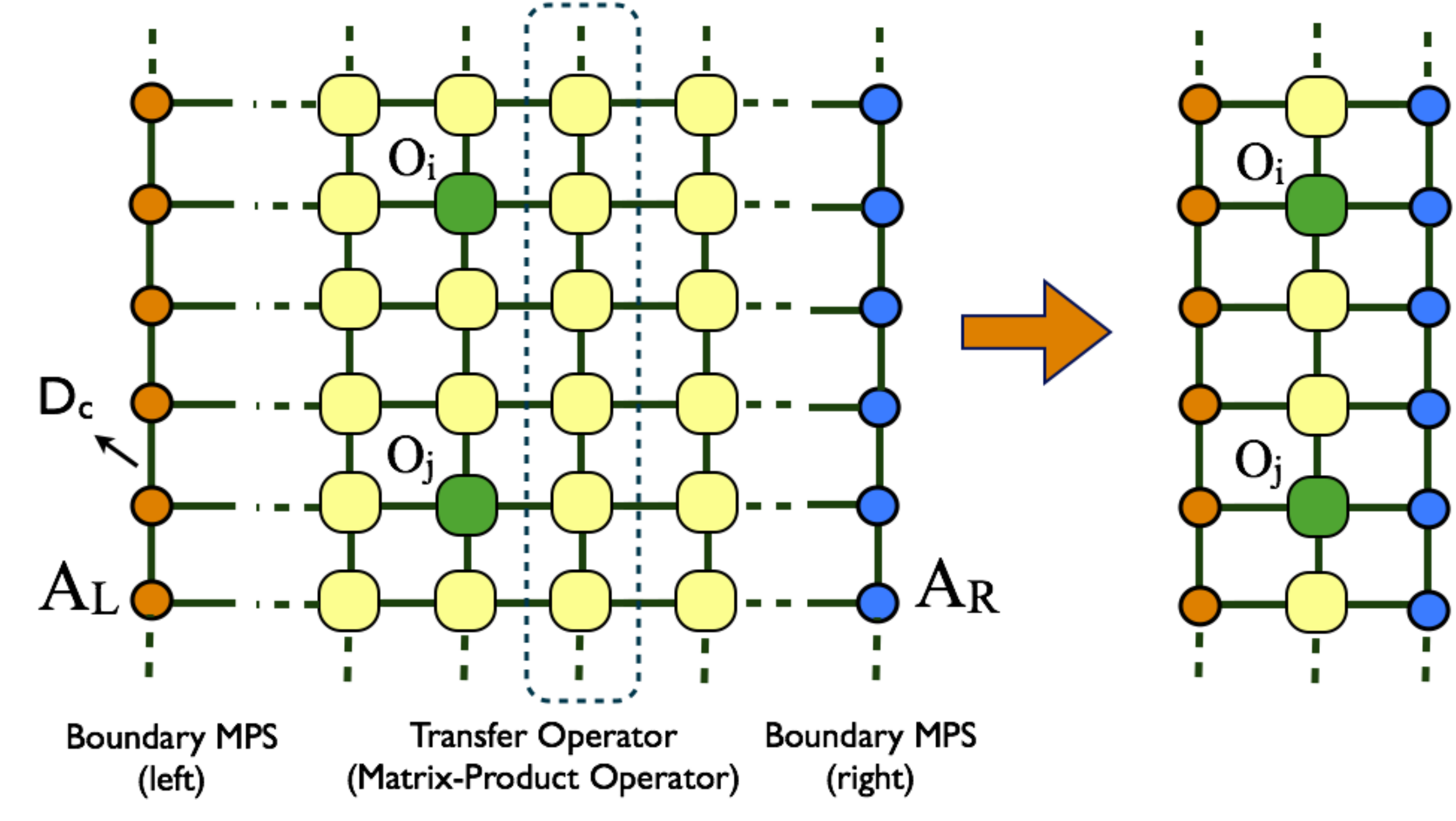}
  \end{center}
  \caption{(Color online) Boundary-MPS contraction scheme in the iPEPS algorithm. The double-layer tensor network is contracted iteratively, by successively contracting the boundary MPS with the transfer operator. The boundary MPS is translationally invariant and includes $A_L$ and $A_R$. The correlation function can be evaluated by sandwiching two (converged) boundary MPS with one column of the transfer operators, including impurity tensors $O_{i(j)}$ denoting the operator at site $i(j)$.}
  \label{fig:iPEPS}
\end{figure}

\subsection{PEPS algorithms}
In our numerical simulations, we mainly deal with the RAL states defined on (i) a cylinder with finite circumference and infinite length and (ii) infinite-size 2D lattices. For the former, when the circumference $L_y$ is relatively small ($L_y \le 5$ for RAL state), we perform exact contractions for extracting physical observables (e.g. correlation functions) and entanglement properties. In the latter case, we employ the infinite-PEPS (iPEPS) algorithm \cite{Jordan-2008, Orus-2009} for approximate contractions with high precision.

Firstly, let us briefly explain the exact contraction method on a cylinder. In order to evaluate the norm of the wave function, or the expectation values of concerned physical quantities, we need to calculate the inner-product of two PEPS wave functions by contracting the double-layer tensor network from both open ends along the (infinitely long) horizontal direction. In each step, the boundary vector is contracted with a single column of transfer operators, and the size of the boundary vectors does not grow up after a step of contraction. Therefore, this process can be repeated until the vectors, as well as the concerned observables, converge. In practice, it takes $10 \sim 20$ iteration steps to converge. However, when treating topological PEPS, we note that the initial boundary vector has to be chosen with special care, since it plays an important role in selecting the topological sectors, as we shall show in later sections.

When both directions of the lattice are infinite, exact contraction is no longer feasible. In this case, we utilize the iPEPS technique \cite{Jordan-2008, Orus-2009} to contract the tensor network with high accuracy. The basic idea is to (approximately) express the boundary vector in the form of a translation-invariant boundary matrix-product state (MPS), and to contract the boundary MPS with the transfer operator (in a matrix-product operator form, see Fig.~\ref{fig:iPEPS}) of the double-layer tensor network. The size of the boundary MPS, i.e., the dimension of its virtual bonds (see Fig.~\ref{fig:iPEPS}), increases exponentially with contraction steps. Therefore, in order to make the contraction process sustainable, it is necessary to truncate the virtual bond space of the boundary MPS after a few steps. This introduces the truncation bond dimension, denoted by $D_c$. We adopt both the standard canonicalization\cite{Vidal-2008} and bi-canonicalization\cite{Xiang-2013} truncation techniques, and compare the results with the extrapolated value obtained in exact contractions to double check the reliability and accuracy of our numerical calculations.

\section{Criticality on bipartite lattices}
In this section, we consider the RAL states on two bipartite lattices, i.e., square and honeycomb lattices. For the honeycomb RAL state [see Fig. \ref{fig:peps} (b) for its PEPS form], we block two nearest-neighbor sites together and thus obtain effectively a square-lattice PEPS. When computing two-point correlation functions, the distances are measured along a line on the effective square lattice, which corresponds to a zig-zag line on the original honeycomb lattice.

\begin{figure}[htpb]
  \begin{center}
	\includegraphics[width=1\columnwidth]{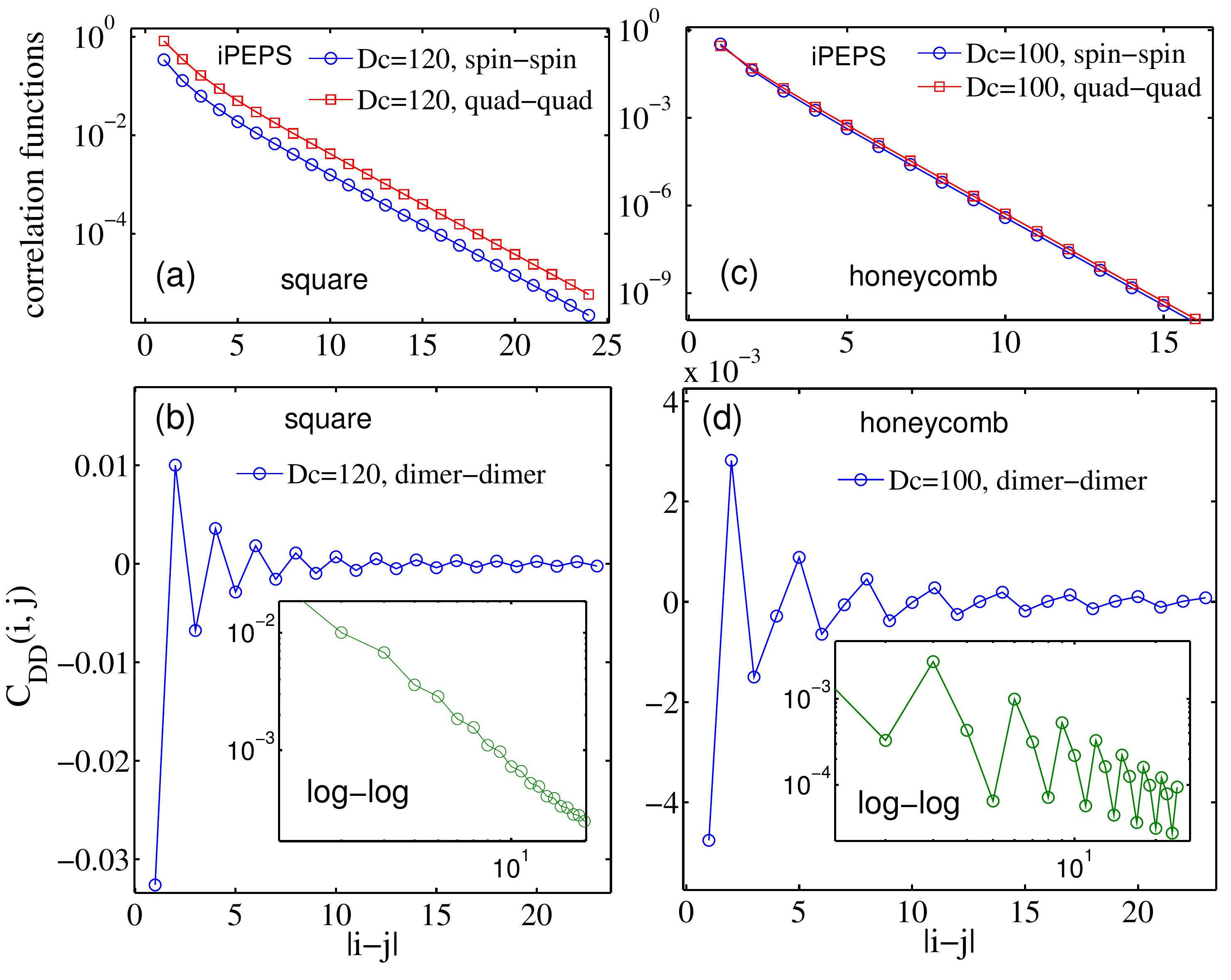}
  \end{center}
  \caption{(Color online) Spin-spin, quadrupole-quadrupole, and dimer-dimer correlation functions of the RAL states on (a,b) square and (c,d) honeycomb lattices. Note the distances $|i-j|$, in both cases, are measured on the square-lattice geometry, and the unit-cell size of square lattice is adopted as the length unit.}
  \label{fig:bipart}
\end{figure}

We employ the iPEPS method to accurately evaluate the correlation functions. When doing this, we have checked the convergence of the results with different truncation parameter $D_c$. The calculated correlation functions are shown in Fig.~\ref{fig:bipart}.

The spin-spin correlation function $C_{\rm{SS}}(i,j)=\langle S^z_i S^z_j \rangle$ and the quadrupole-quadrupole correlation function $C_{\rm{QQ}}(i, j) = \langle Q^z_i Q^z_j \rangle$, where $Q^z = \frac{1}{\sqrt{3}} [3(S^z)^2-2]$, both decay exponentially as can be seen in Fig.~\ref{fig:bipart}(a,c), suggesting the absence of N\'{e}el or quadrupolar long-range order. However, the dimer-dimer correlation function, $C_{\rm{DD}}(i, j) = \langle (\bold{S}_i \cdot \bold{S}_{i+1}) (\bold{S}_j \cdot \bold{S}_{j+1}) \rangle - \langle \bold{S}_i  \cdot \bold{S}_{i+1} \rangle \langle \bold{S}_j \cdot \bold{S}_{j+1} \rangle$, decays algebraically on both lattices, as shown in Fig. \ref{fig:bipart}(b,d). Therefore, we conclude that the RAL states are critical on square and honeycomb lattices~\cite{HYao-2010}. We expect that it is a consequence of the fully-packed arrangement of the loops, similar to those critical loop models in classicial statistical mechanics~\cite{Nienhuis-1982, Nienhuis-1994}.

\section{The RAL states on a kagome lattice}

\subsection{$\mathbb{Z}_2$ topological order}
\label{sec: RAL-topo}
Now we turn to the non-bipartite kagome lattice and provide numerical evidence that the kagome RAL state is gapped and possesses $\mathbb{Z}_2$ topological order. In this Section, we only present results of the RAL state with sign convention shown in Fig. \ref{fig:peps}(c). For the other sign convention as in Fig. \ref{fig:peps}(d), the results lead to the same conclusion and thus are not present here.

\begin{figure}[htpb]
  \begin{center}
	\includegraphics[width=1\columnwidth]{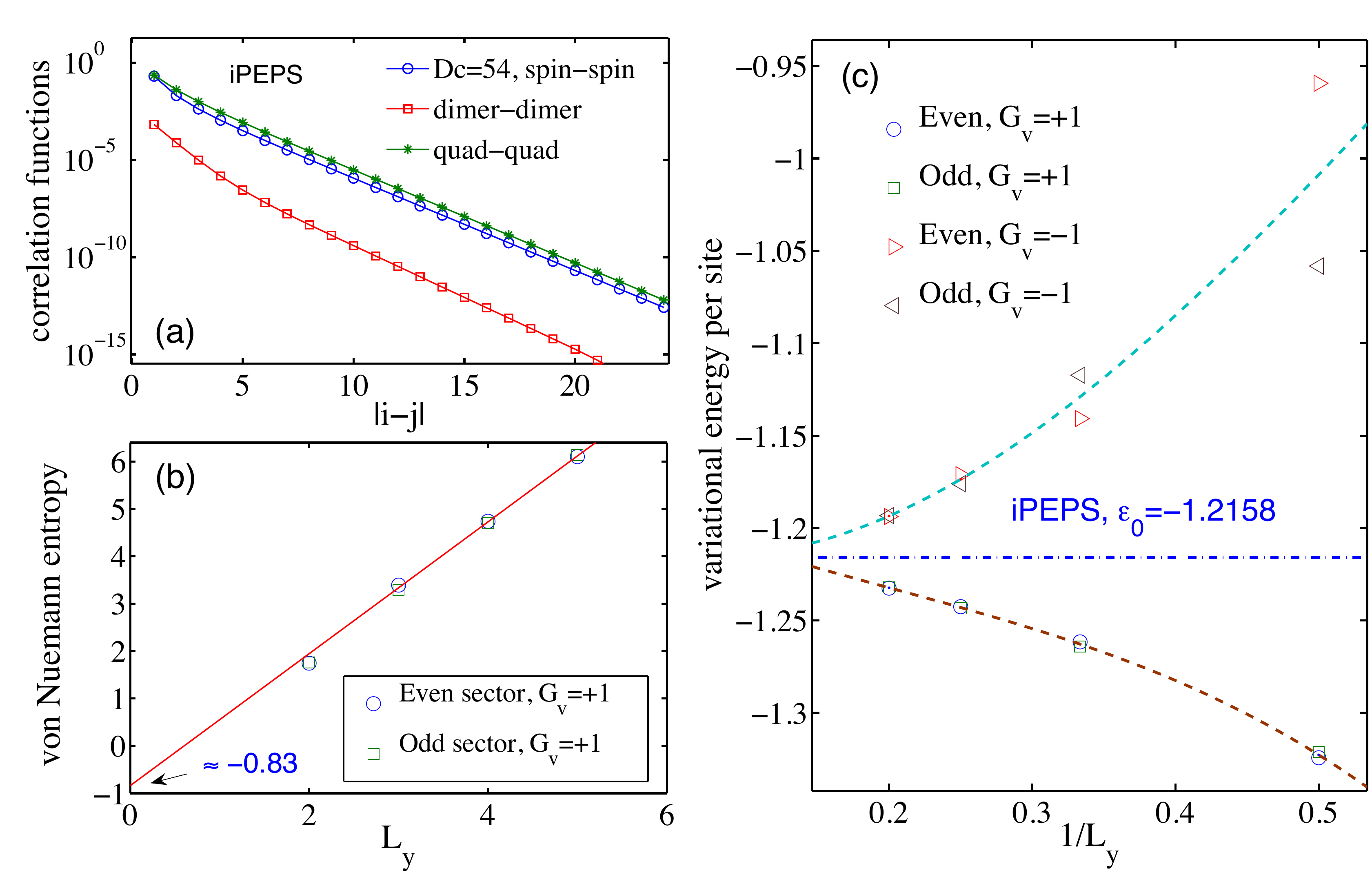}
  \end{center}
  \caption{(Color online) (a) Spin-spin, dimer-dimer, and quadrupole-quadrupole correlation functions of the RAL state on the kagome lattice (measured on the square-lattice geometry). (b) The von-Neumann entropy $S(L_y)$ of the kagome RAL state in the two topological sectors with $G_v=1$. $S(L_y)$ for circumferences $L_y=3,4,5$ are used in the fit. (c) The energy expectation value (estimated from NN spin-spin correlation function) of four topological sectors, all of which draw near the iPEPS result when $L_y$ increases. The sign convention of the RAL state is given in Fig. \ref{fig:peps}(d).}
  \label{fig:kagome}
\end{figure}

We first compute the spin-spin, dimer-dimer and quadrupole-quadrupole correlation functions. These results are shown in Fig.~\ref{fig:kagome}(a). We note that, by adopting the simplex construction in Fig.~\ref{fig:peps-kag}(b), one obtains an effective honeycomb-lattice representation of the RAL states, which can then be transformed to a square-lattice PEPS as we did in the previous Section. The correlation functions are thus measured along a line of this coarse-grained square lattice. The correlation functions in Fig.~\ref{fig:kagome}(a) all exhibit exponentially decaying behaviors, supporting that the kagome RAL state is a gapped spin liquid.

To characterize the RAL state on the kagome lattice, we compute the entanglement entropy on an infinite cylinder following the scheme described in Ref. [\onlinecite{Jiang-2012}], utilizing the PEPS technique in Ref. [\onlinecite{Cirac-2011}]. The cylinder is assumed to be periodic (open) along the vertical (horizontal) direction. We trace over the states on half of the cylinder to obtain the reduced density matrix and the entanglement entropy. Before presenting the numerical results, we first analyze the topological sectors on the cylinder to construct the so-called minimally entangled states~\cite{YZhang-2012, YZhang-2011,Poilblanc-2012, Schuch-2012, Schuch-2013, Poilblanc-2013}. Since the wave function closely resembles the string picture of a $\mathbb{Z}_2$ topological phase (or a $\mathbb{Z}_2$ gauge theory), we use the language of the $\mathbb{Z}_2$ gauge theory throughout our discussion.

\begin{figure}[htpb]
  \begin{center}
	\includegraphics[width=0.92\columnwidth]{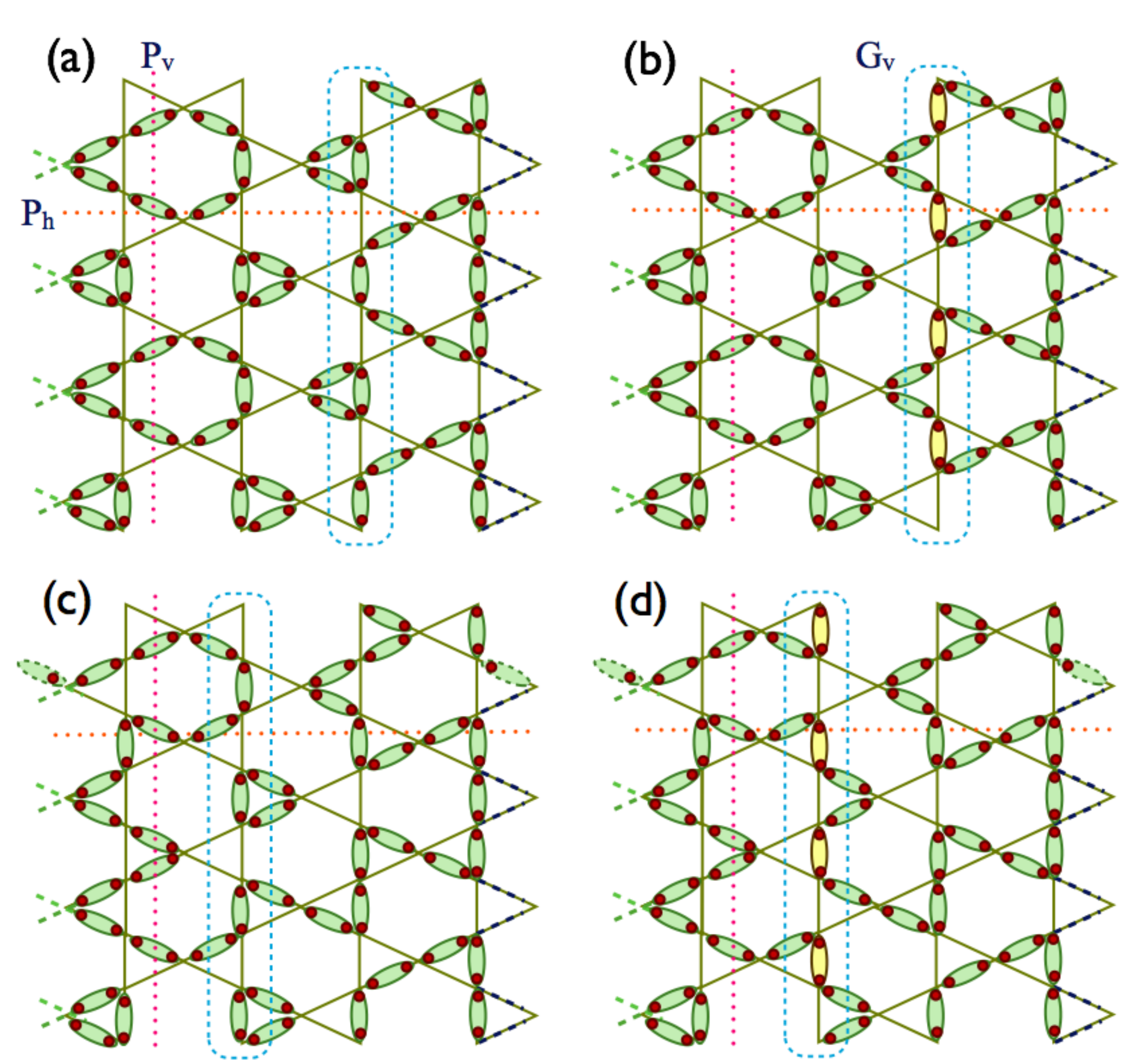}
  \end{center}
  \caption{(Color online) Illustrations of topological sectors of the kagome RAL state. (a) $P_h=P_v=1$. (b) $P_h=-1, P_v=1$. (c) $P_h=1, P_v=-1$. (d) $P_h=P_v=-1$. Vertical (horizontal) direction of the cylinder is periodic (open). $G_v$ means shifting the valence bonds along the vertical direction, which is a global move changing the winding number of the loops around the cylinder circumference and thus the parity of $P_h$.}
  \label{fig:topo_sect}
\end{figure}

\begin{figure}[htpb]
  \begin{center}
	\includegraphics[width=0.7\columnwidth]{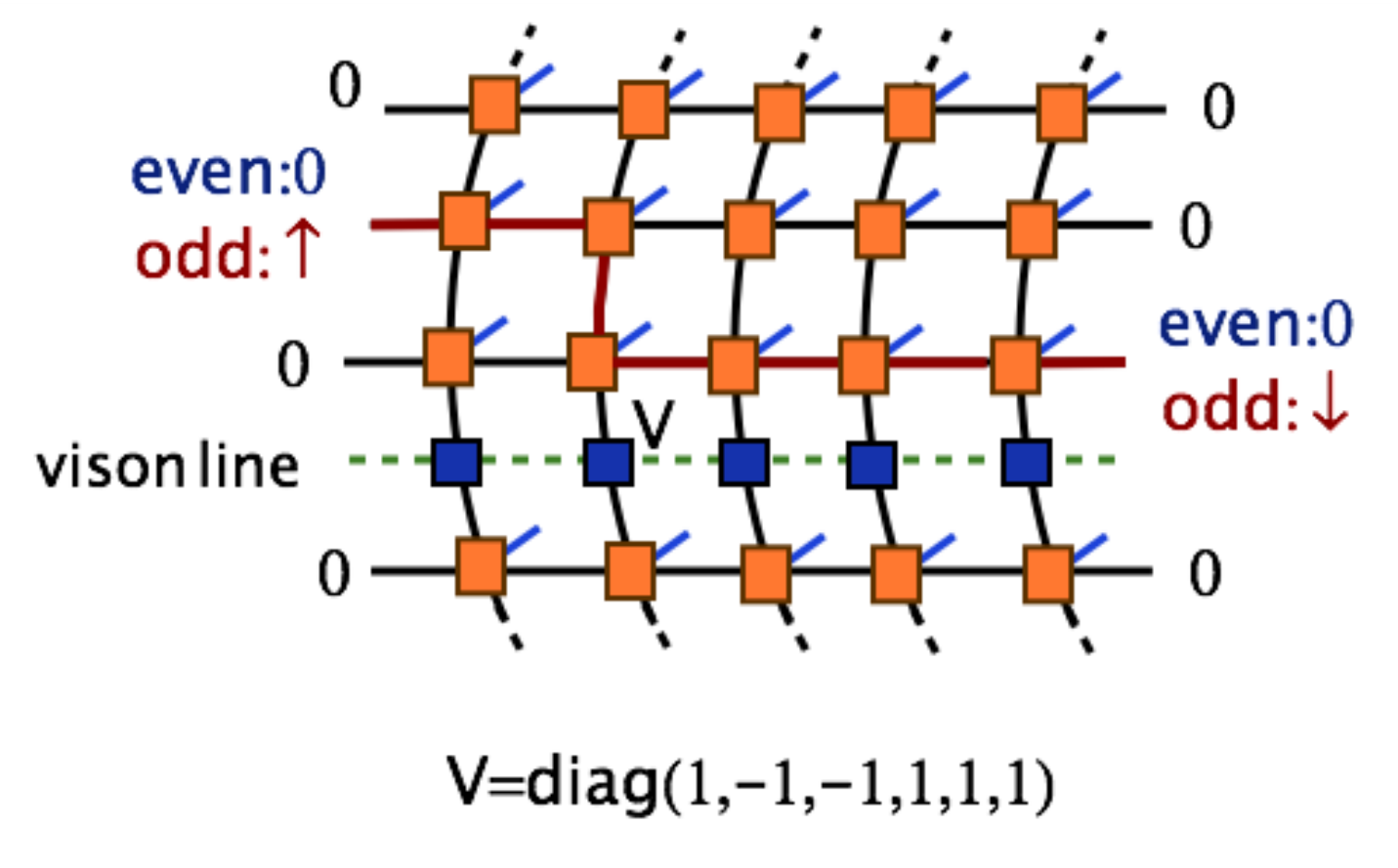}
  \end{center}
  \caption{(Color online) By threading a magnetic flux (vision line) and/or introducing an open electric field line (connecting the two boundary spinons) along the cylinder, we can generate four topological sectors ($G_v=\pm1, P_v=\pm1$) of the kagome RAL states. The dashed line consists of diagonal $v$ matrices is a vision line. $\uparrow$ ($\downarrow$) on the boundary is spinon with up (down) spin.}
  \label{fig:cylinder}
\end{figure}

We can easily identify four topological sectors distinguished by two non-local operators, the magnetic flux along the horizontal (vertical) direction denoted by $P_{h(v)}$, which counts the parity of the number of AKLT valence bonds along a horizontal (vertical) cut. First we set both ends of the cylinder open (i.e. AKLT lines can not terminate on the ends). The two sectors can be chosen as eigenstates of $P_h$ with eigenvalues $\pm 1$,  as shown in Fig. \ref{fig:topo_sect}(a,b). Shifting the virtual valence bonds around the circumference of the cylinder toggles between the two sectors. This operator is exactly the Wilson loop denoted by $G_v$, i.e., create a pair of spinons by breaking a virtual valence bond, wind one spinon out of the two around the cylinder and annihilate them again on the other side). In practice, we insert a line of diagonal $v$-matrices to the PEPS (i.e., thread a magnetic flux along the cylinder, see Fig.~\ref{fig:cylinder} for illustration), and then construct two eigenstates of $G_v$: $ \ket{G_v =\pm1} = \frac{1}{\sqrt{2}}(\ket{P_h=1}\pm \ket{P_h=-1})$ which are minimally entangled~\cite{YZhang-2012} and ready for entanglement entropy calculation.

In order to construct the other two sectors, we are enforced to introduce virtual polarized spin-$1/2$ spinons [between which there is an open string of AKLT state, see Fig. \ref{fig:topo_sect}(c,d)] on the two boundaries of the cylinder. They act as two $\mathbb{Z}_2$ charges where the electric field lines can terminate. It is easy to see that a vertical path necessarily cuts through odd number of virtual spin-1/2 valence bonds with this particular boundary conditions~\cite{comment:sym-breaking}.  We refer to the sectors with this boundary condition as the odd topological sector ($P_v=-1$), while the previous two with open boundaries as the even ones ($P_v=1$). In the even sector, the reduced density matrix only has nonzero eigenvalues corresponding to eigenvectors with integer total spin; while in the odd sector, because of the virtual spin-1/2 spinons on the boundaries, all Schmidt eigenstates have half-odd-integer total spin, i.e., each level in the entanglement spectrum has an even-fold degeneracy (a detailed analysis on entanglement spectra will be present latter in section \ref{sec:ES}), which we have confirmed numerically~\cite{comment:sym-breaking}. This is a manifestation of the fact that the $\mathbb{Z}_2$ charge forms a  $S=\frac{1}{2}$ representation of the $\mathbb{SO}(3)$ symmetry group while the $\mathbb{Z}_2$ flux excitation carries no nontrivial projective representation (i.e. integer spin).

In Fig.~\ref{fig:kagome}(b), we present the entanglement entropies of the two minimally entangled states $\ket{G_v=1}$ in even and odd sectors. The entanglement entropy $S(L_y)$ as a function of the circumference $L_y$ can be evaluated using both converged (left and right) boundary vectors obtained from the exact contraction \cite{Cirac-2011}. The topological entropy $\gamma$ is extracted by fitting $S(L_y)$ to $S(L_y)=\alpha L_y-\gamma$ and we find $\gamma\approx 0.83$ (close to $\ln2$) in both sectors, indicating that the kagome RAL state has {a} $\mathbb{Z}_2$ topological order.

Fig.~\ref{fig:kagome}(c) shows the variational energies of the RAL states in all four topological sectors with respect to the spin-1 antiferromagnetic Heisenberg model on the kagome lattice. When increasing $L_y$, the variational energies of the wave functions in four different topological sectors ($G_{v}=\pm1$, $P_v=\pm1$) converge rapidly, all getting close to the value $-1.2158$ obtained by iPEPS method, suggesting that they are degenerate in the thermodynamic limit.

\begin{figure}[htpb]
  \begin{center}
	\includegraphics[width=1\columnwidth]{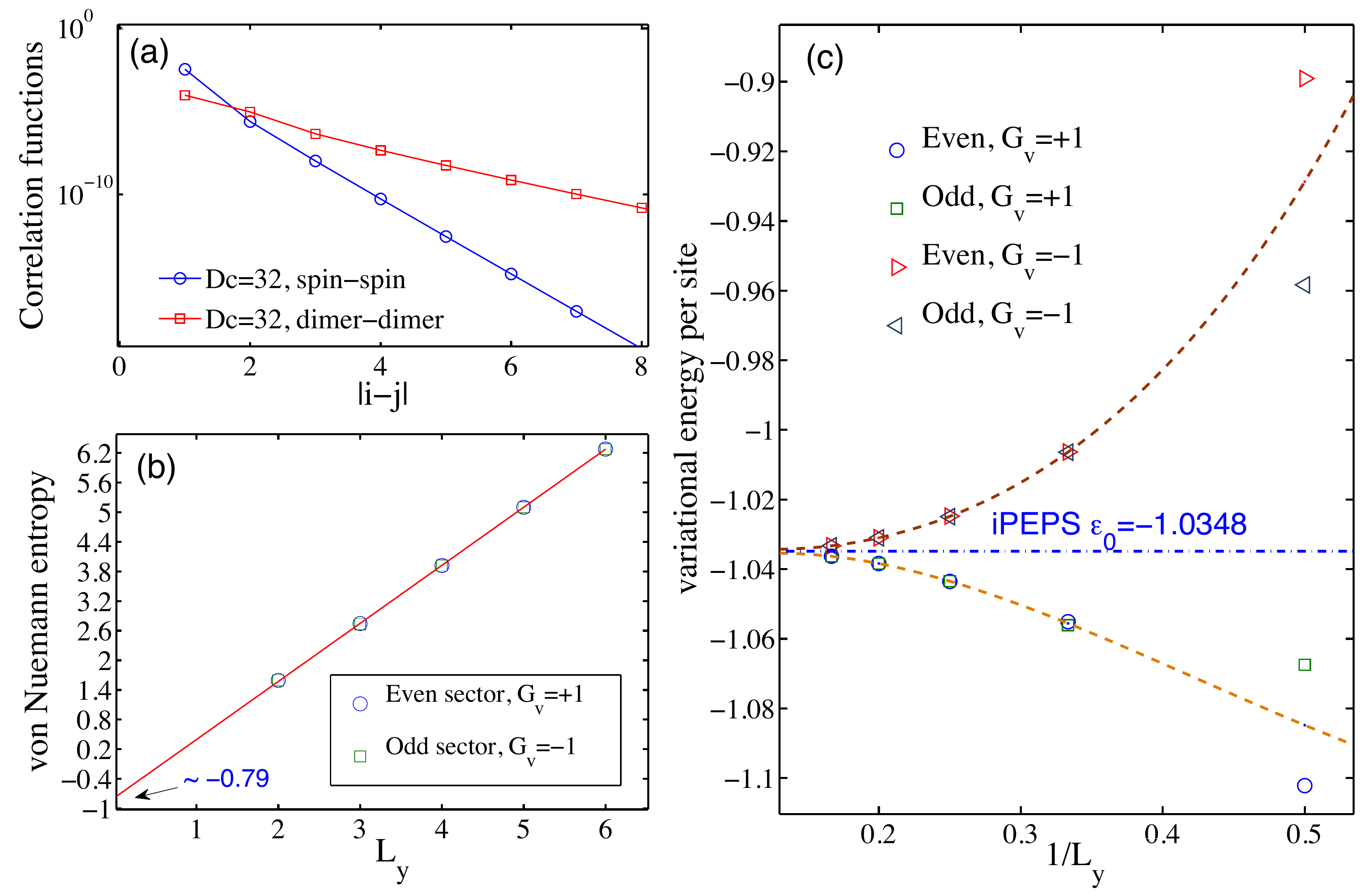}
  \end{center}
  \caption{(Color online) (a) The spin-spin and dimer-dimer correlation functions of spin-1 RVB state, both of which decay exponentially. (b) The variational energies of four topological sectors, all of which converge rapidly to the iPEPS result in the thermodynamic limit. (c) Entanglement entropies of the spin-1 RVB state in the topological sectors with $G_v=1$. By extrapolating the data the topological entanglement entropy is found to be $\gamma = 0.79$, close to $\ln 2$.}
  \label{fig:spin1-RVB}
\end{figure}

\subsection{Characterization of the spin-1 RVB state}
\label{sec: spin1-RVB}
So far we have constrained ourselves to the pure RAL state ($\alpha=0$ in Eq. \ref{eq:interpolation}), and excluded the spin-1 valence bonds (equivalent to a short AKLT loop with length 2). However, it is interesting to consider the other limit ($\alpha=1$) where only the spin-1 short-range valence bonds are allowed.

In the case of $\alpha=1$, we have a spin-1 short-range RVB state, which can be represented by a PEPS with bond dimension $D=4$, i.e, the virtual spin is in the representation $0 \oplus 1$. The vision line can be defined as $v = \rm{diag}(1, -1, -1, -1)$, and the odd topological sector is constructed by attaching a pair of virtual spin-1 on both ends.

In Fig.~\ref{fig:spin1-RVB}, we present the correlation functions (including the variational energies of the kagome Heisenberg model from nearest-neighbor spin-spin correlations) and the entanglement entropy of the spin-1 RVB state on a kagome lattice. Fig.~\ref{fig:spin1-RVB}(a) shows that both the spin-spin and dimer-dimer correlation functions decay exponentially, suggesting that the spin-1 kagome RVB state is fully gapped. In Fig.~\ref{fig:spin1-RVB}(b), we calculate the variational energies of the spin-1 kagome Heisenberg model in four topological sectors. All of them converge to the iPEPS energy result rapidly when the circumference increases. Fig.~\ref{fig:spin1-RVB}(c) shows the von Neumann entanglement entropies as a function of the circumference $L_y$, from which we extract the topological entanglement entropy $\gamma \approx 0.79$ (close to $\ln 2$). This is a clear indication of the underlying $\mathbb{Z}_2$ topological order.

\subsection{Interpolating between the RAL and RVB states}
In the previous subsections, we have shown that both the pure RAL ($\alpha=0$) and the spin-1 RVB ($\alpha=1$) states on a kagome lattice have $\mathbb{Z}_2$ topological order. As mentioned in Sec. \ref{sec:kagomeRAL}, the $D=6$ PEPS defined by Eq.~\ref{eq:interpolation} can describe the states with mixed loop and dimer configurations on equal footing, and we can tune the parameter $\alpha$ to interpolate between the pure RAL and spin-1 RVB states.

\begin{figure}[htpb]
  \begin{center}
	\includegraphics[width=0.9\columnwidth]{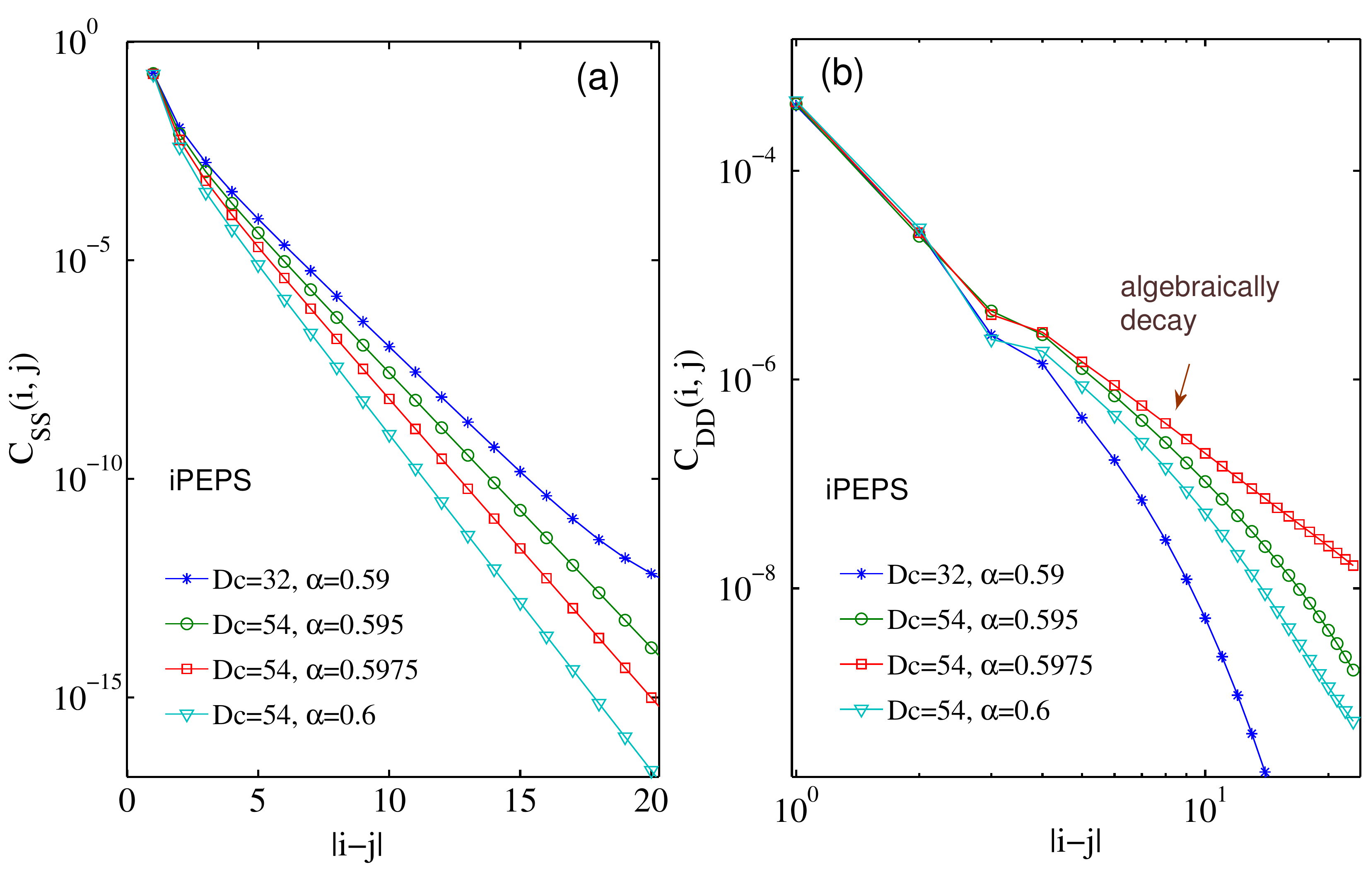}
  \end{center}
  \caption{(Color online) Correlation functions of the mixed RAL state on kagome lattice. (a) The spin-spin correlation functions $C_{\rm{SS}}(i, j)$ always decay exponentially for various $\alpha$ values. (b) At $\alpha_c \approx 0.5975$, there is a gapless point with algebraic dimer-dimer correlation function $C_{\rm{DD}}(i,j)$.}
  \label{fig:gapless}
\end{figure}

\begin{figure*}[htpb]
\begin{center}
(a)\includegraphics[width=0.75\columnwidth]{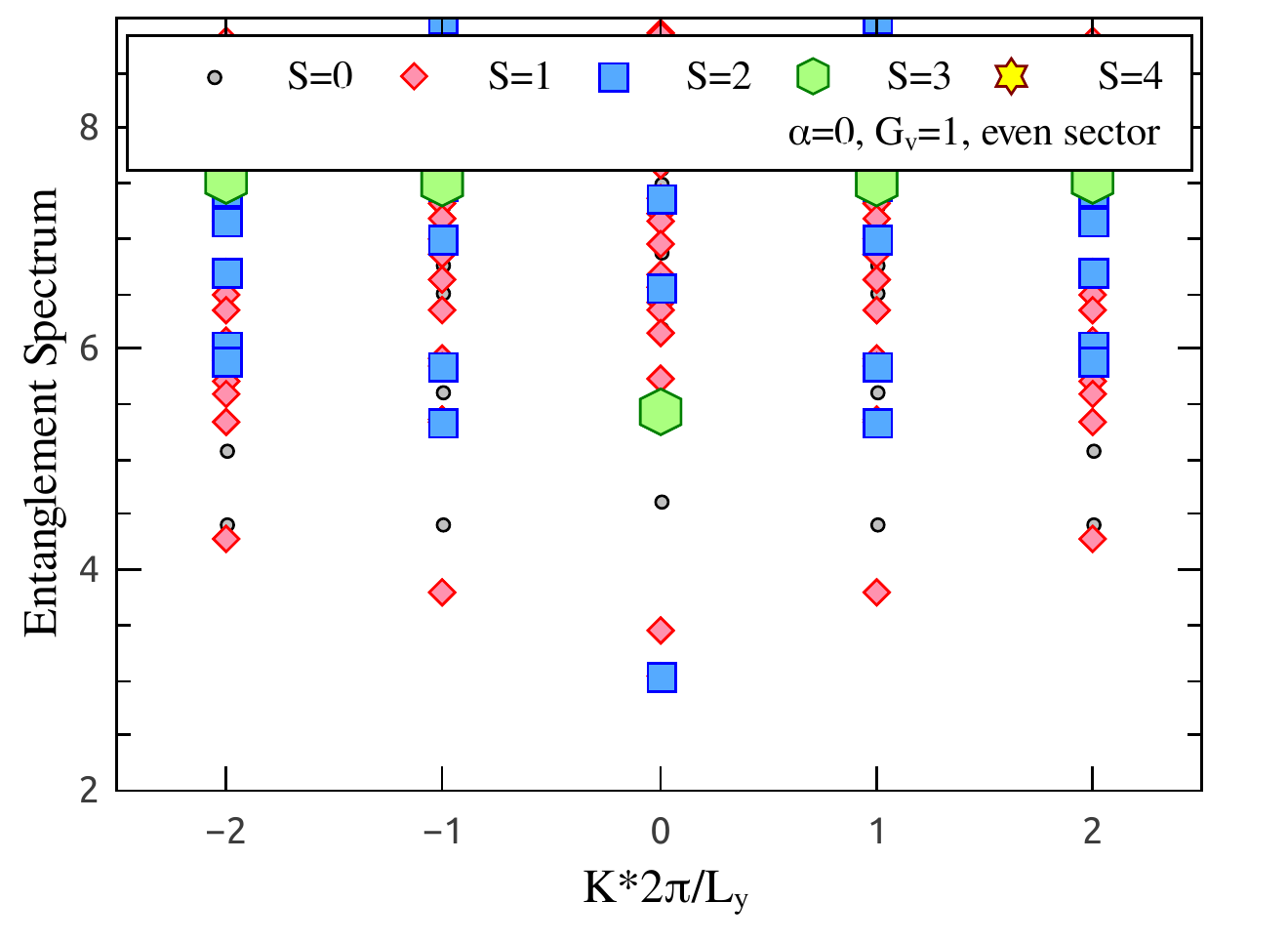}
(b)\includegraphics[width=0.75\columnwidth]{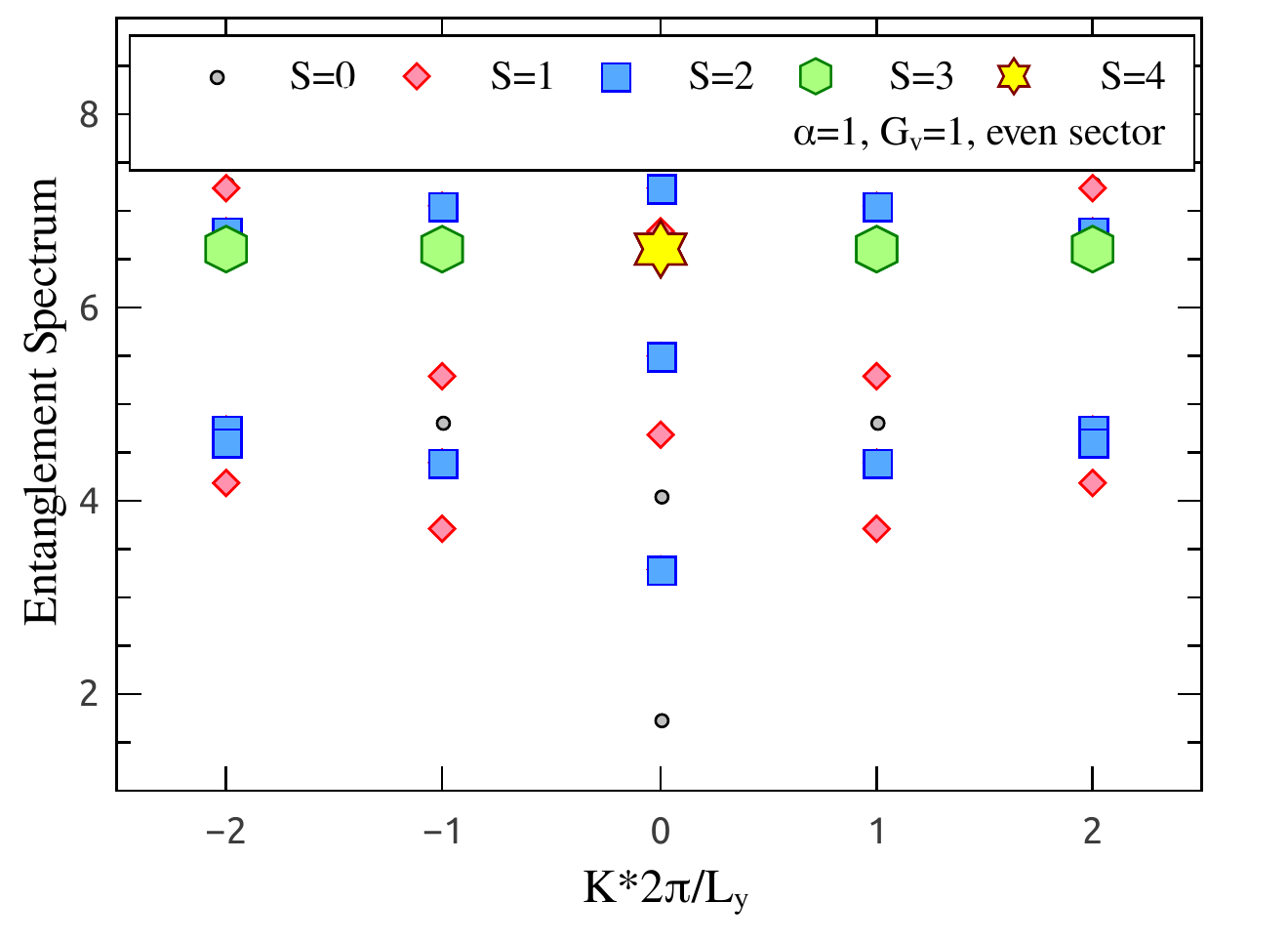}
(c)\includegraphics[width=0.75\columnwidth]{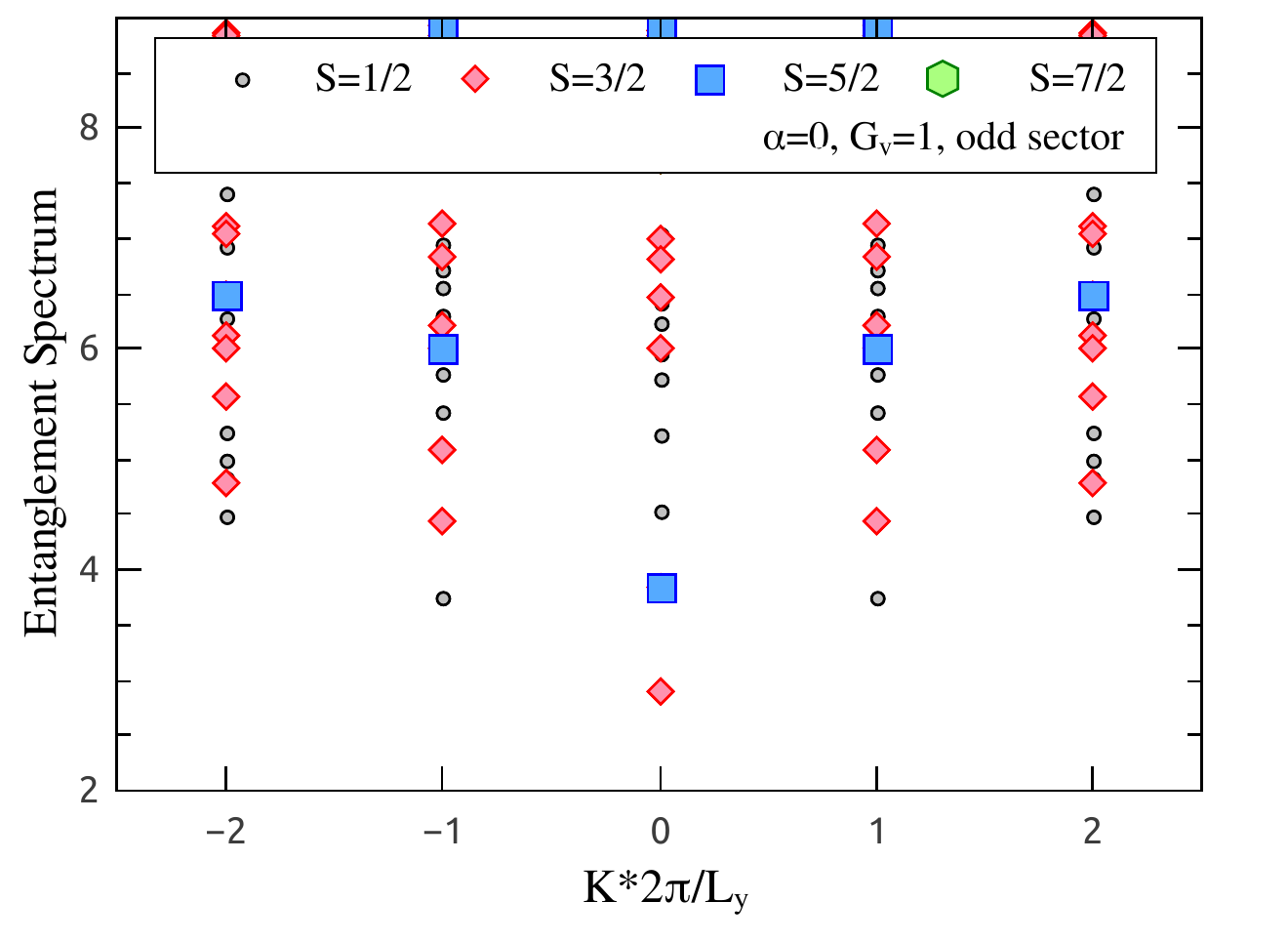}
(d)\includegraphics[width=0.75\columnwidth]{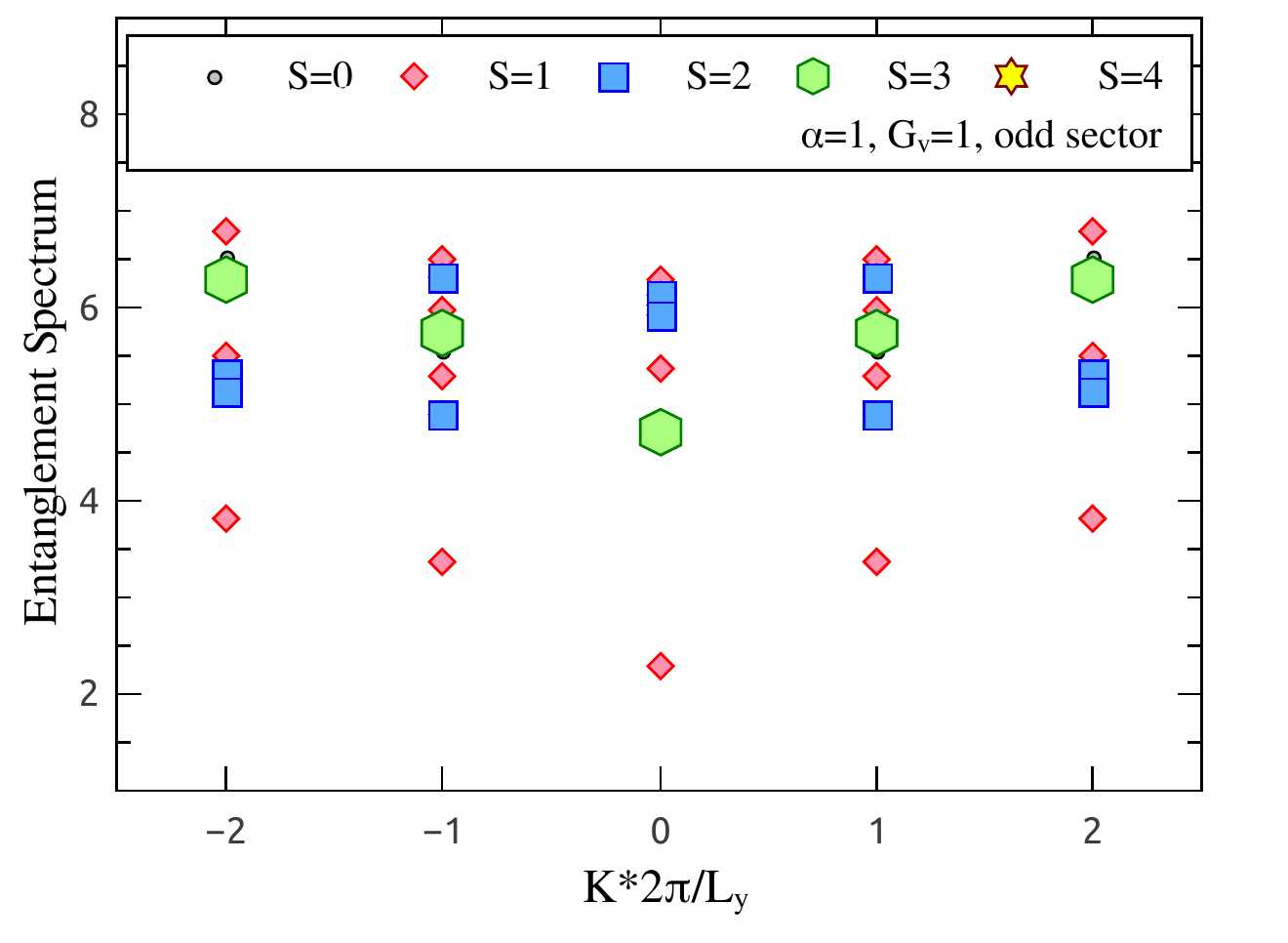}
\end{center}
\caption{(Color online) Entanglement spectrum versus the total momentum for an infinitely long kagome cylinder with circumference $L_y=4$. (a) $\alpha=0$ (RAL), even sector. (b) $\alpha=1$ (RVB), even sector. (c) $\alpha=0$ (RAL), odd sector. (d) $\alpha=1$ (RVB), odd sector. Each point in the spectrum represents a spin-$S$ multiplet with $2S+1$ individual states. Only the low-lying levels of the spectra are shown.}
\label{fig:ES-momentum}
\end{figure*}

The following question arises naturally: can this path adiabatically connect the spin-1 RAL and RVB states? Theoretically, although the two limiting states have identical intrinsic topological order, they behave very differently under $\mathbb{SO}(3)$ spin symmetry. To be more concrete, the RVB state has no deconfined half-integer spin excitations. Since the path under consideration preserves the $\mathbb{SO}(3)$ symmetry all the way from $\alpha=0$ to $\alpha=1$, we expect there must be at least one singular point where the state becomes critical. In Fig. \ref{fig:gapless} we show that, although the spin-spin correlators always decay exponentially [see Fig. \ref{fig:gapless}(a)], the dimer-dimer correlation function decays algebraically at around $\alpha_c \approx 0.5975$, as depicted in Fig. \ref{fig:gapless}(b).

For this one-parameter family of PEPS $\ket{\Psi_\alpha}$, one can always find a local, positive semidefinite Hamiltonian $H(\alpha)$ such that $H(\alpha) \ket{\Psi_\alpha} =0$, i.e., the RAL state $\ket{\Psi_\alpha}$ is the exact ground state of this co-called parent Hamiltonian~\cite{Schuch-2010}. The existence of a critical point implies that the path in the space of parent Hamiltonians should be gapless at $\alpha_c$. However, as the dimension of virtual Hilbert space for this PEPS is $D=6$, one needs to block a large number of sites so that a local projector annihilating the PEPS exists, which constitutes the parent Hamiltonian. Due to its complicated form, we do not present the analytical form of the parent Hamiltonian $H(\alpha)$ here.

\subsection{Entanglement spectra}
\label{sec:ES}

\begin{figure}[htpb]
  \begin{center}
	\includegraphics[width=0.85\columnwidth]{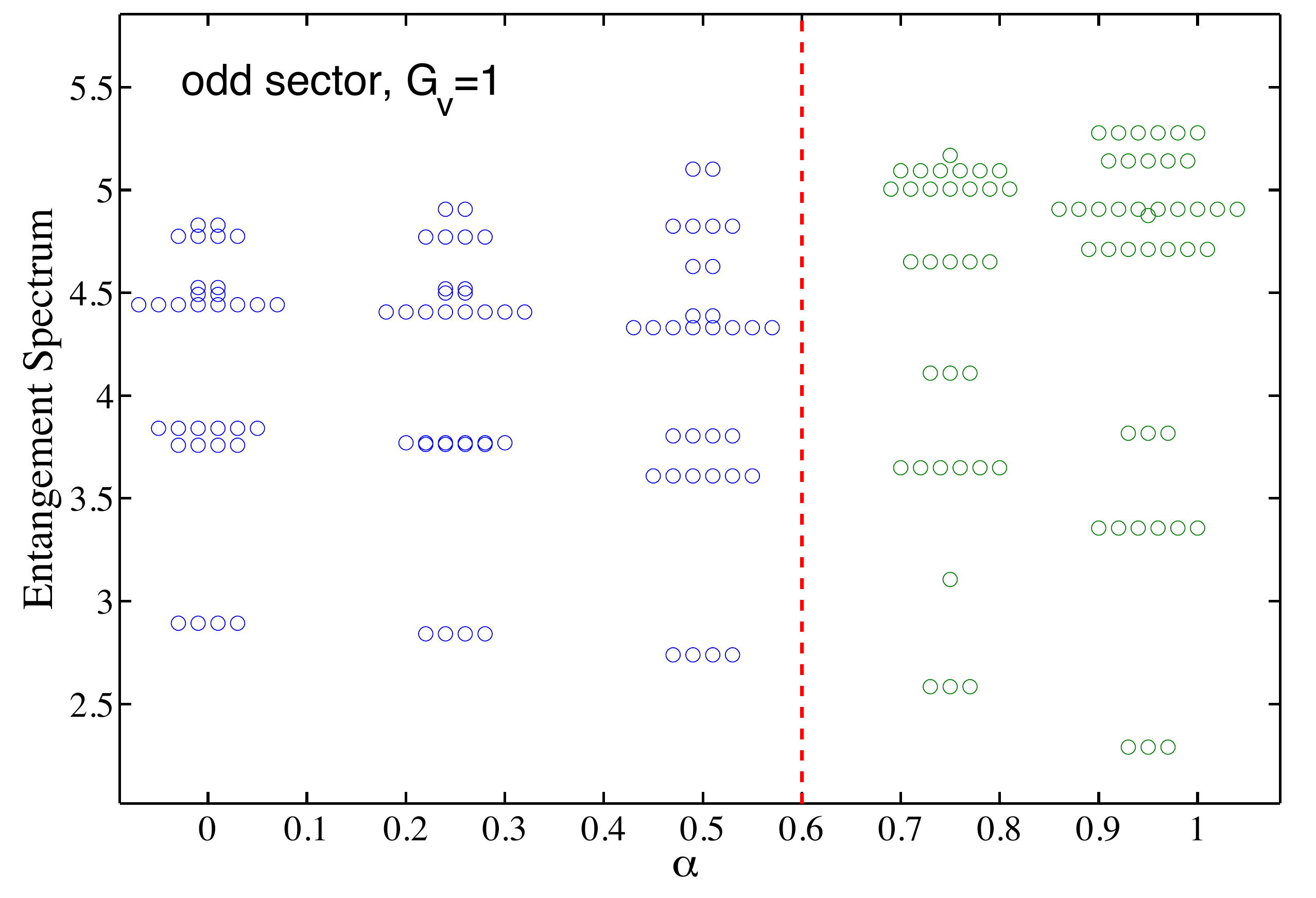}
  \end{center}
  \caption{(Color online)  Entanglement spectra for the mixed RAL states. A pair of virtual spinons are put on the open ends of the cylinder. Each level in the spectrum has even-fold degeneracy for $\alpha < 0.6$ (RAL region, left-hand side of the red dashed line); while for $\alpha > 0.6$ (RVB region on the right side) they do not have this feature in the level degeneracies. The spectra are evaluated on infinte-long cylinders with circumference $L_y=4$.}
  \label{fig:ES}
\end{figure}

Entanglement spectrum (ES) is defined as the eigenvalues of the operator $-\ln \rho_R$, where $\rho_R$ is the reduced density matrix. The full ES has been suggested to reveal entanglement properties, providing more useful information than the entanglement entropy~\cite{Haldane-2008}.

In this section, we provide the entanglement spectra of the pure RAL, spin-1 RVB and the mixed RAL states in different topological sectors. The construction of four topological sectors for the pure RAL and spin-1 RVB states have been discussed in Secs. \ref{sec: RAL-topo} and \ref{sec: spin1-RVB}.

First we consider the ES of the pure RAL ($\alpha=0$) and spin-1 RVB ($\alpha=1$) states, shown in Fig.~\ref{fig:ES-momentum}. We have taken advantange of the $\mathbb{SO}(3)$ spin-rotational symmetry and the translation symmetry in the vertical (circumference) direction of the cylinder and label each level in the spectrum by its spin $S$ (a spin-$S$ multiplet represents $2S+1$ states) and momentum $K$ (in vertical direction). For both states, the ES in the even sectors are filled with integer-spin multiplets and do not have any recognizable features regarding the degeneracy. However, the RAL and RVB states have distinct ES in the odd sectors. The ES levels of the RAL state all have half-integer spins, giving rise to even-fold degeneracies in the spectrum. For the spin-1 RVB state, the ES in the odd sector are still filled with integer-spin multiplets.

One interesting question regarding the entanglement spectrum shown in Fig.~\ref{fig:ES-momentum} is whether the ``low energy" part of the ES is critical and described by a conformal field theory. However, as the largest system size we can achieve for computing the ES is $N_y=5$, which is still too small for a finite-size scaling of the gap, we do not have a definite answer to this question and leave it for a future study.

For the mixed RAL states with $0<\alpha<1$, it is not entirely clear how to construct the topological sectors. Due to the phase transition at $\alpha_c$, we expect that the ES for $\alpha<\alpha_c$ in the odd sector is qualitatively different from those for $\alpha>\alpha_c$. Since the mixed RAL state for $\alpha < \alpha_c$ is adiabatically connected to the pure RAL state, we expect that the four topological sectors also evolve adiabatically in this parameter region. Formally we can construct a series of wave functions, for both $\alpha<\alpha_c$ and $\alpha>\alpha_c$, with the same boundary conditions (i.e. put a pair of oppositely polarized spinons on two open ends of the cylinder \cite{comment:sym-breaking}). We calculate the ES of these states and the results are shown in Fig. \ref{fig:ES}. All levels in the ES have even-fold degeneracies for $\alpha < \alpha_c$ (RAL region), while they are featureless regarding the degeneracy for $\alpha > \alpha_c$ (RVB region).

\section{Variational study of the spin-1 kagome Heisenberg model}
In this section, we use the RAL states as variational wave functions for the spin-1 Heisenberg model
\begin{equation}
H= \sum_{\langle i,j\rangle} \mathbf{S}_i \cdot \mathbf{S}_j
\end{equation}
on the kagome lattice, whose ground-state properties remain elusive to the best of our knowledge.

\begin{figure}[htpb]
  \begin{center}
	\includegraphics[width=0.7\columnwidth]{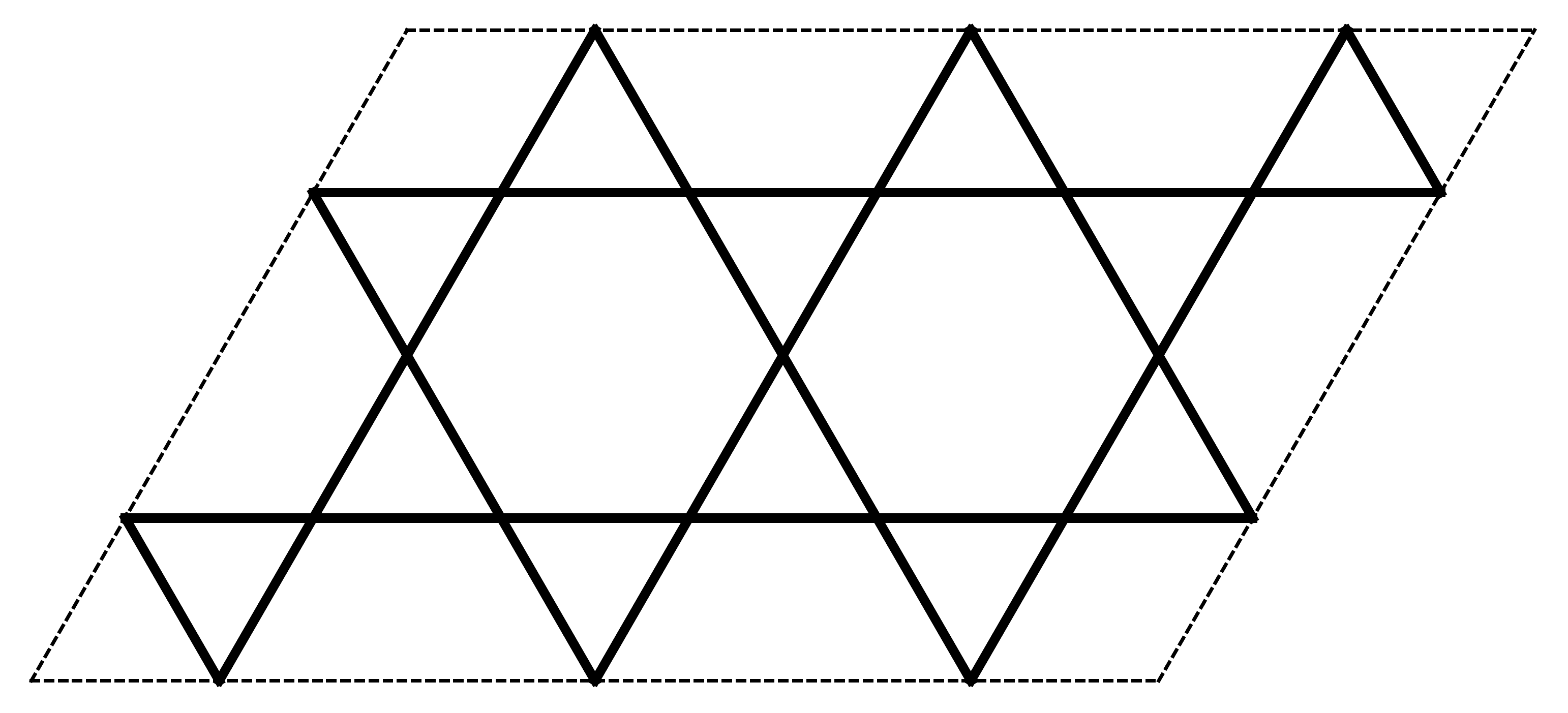}
  \end{center}
  \caption{The 18-site cluster used in the variational study of spin-1 kagome Heisenberg model. Periodic boundary conditions are assumed in both directions, and the cluster has $2\times3$ unit cells.}
  \label{fig:torus}
\end{figure}

We start with the pure RAL state as a parameter-free variational wave function for spin-1 kagome Heisenberg model. We first compare with the exact diagonalization (ED) result on a 18-site torus with $2 \times 3$ unit cells (see Fig. \ref{fig:torus}), which is the largest size considered in the ED study\cite{Hida}. We place the PEPS tensors of RAL states on this small torus geometry, and evaluate the energy expectation value by exact contraction. The variational energy of the pure RAL state with the sign convention given in Fig. \ref{fig:peps}(c) is $-1.383$ per site, closer to the ED value $-1.4393$ than that of the hexagonal singlet solid variational state ($\simeq  -1.304$ per site)~\cite{Hida, Hida_HSS}.

\begin{figure}[htpb]
  \begin{center}
	\includegraphics[width=0.9\columnwidth]{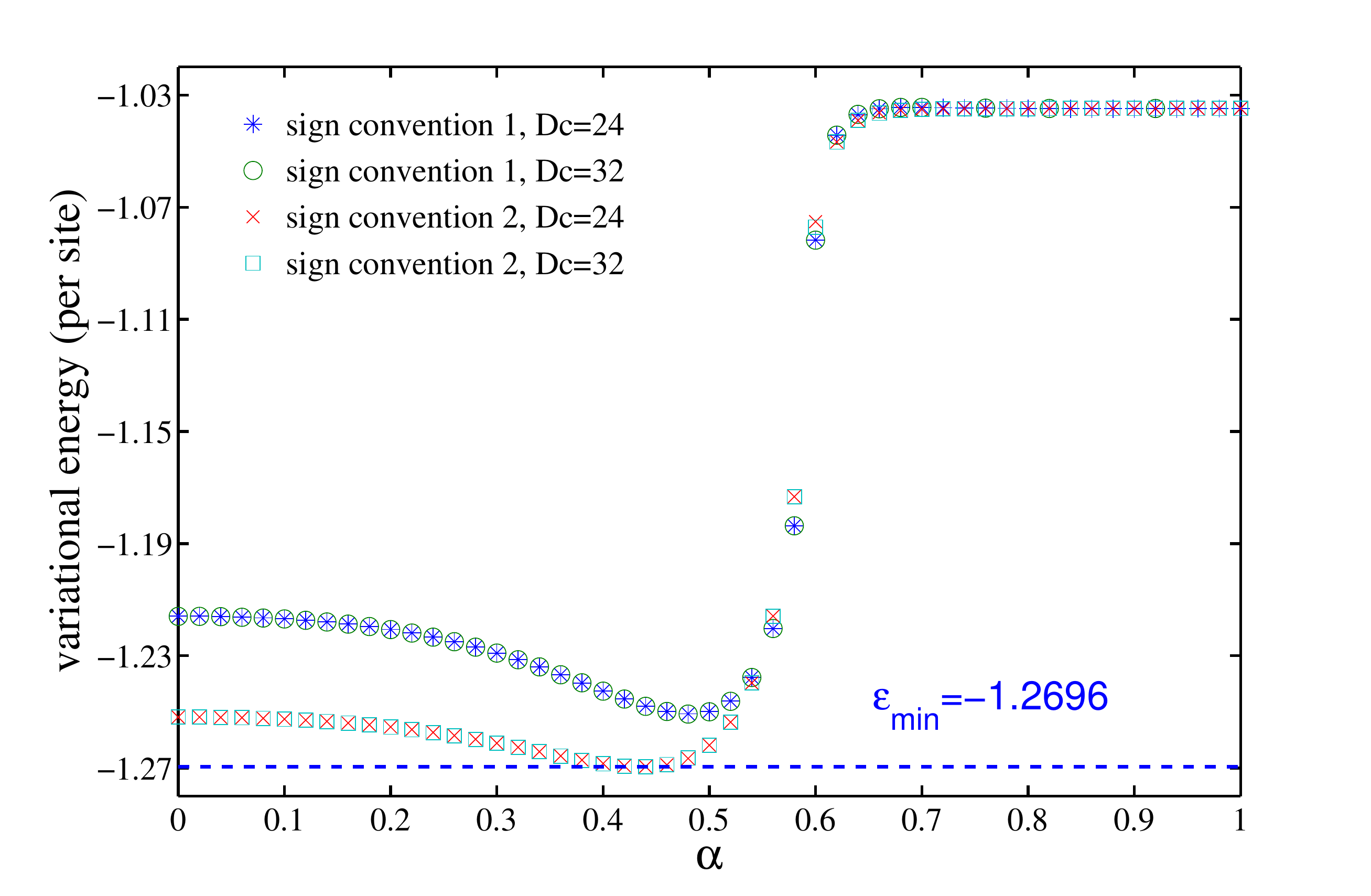}
  \end{center}
  \caption{(Color online) Variational energy of the mixed RAL states for spin-1 kagome Heisenberg model, computed using iPEPS. Convergence has been checked against different $D_c$. Sign convention 1 is illustrated in Fig. \ref{fig:peps}(c) and convention 2 in Fig. \ref{fig:peps}(d).}
  \label{fig:vari-eg}
\end{figure}
Next, we extend the variational study to the one-parameter family of PEPS wave functions, i.e., the mixed RAL states, which interpolate the pure RAL and spin-1 RVB states (controlled by the parameter $\alpha$). The results from the iPEPS calculations in the thermodynamic limit is shown in Fig.~\ref{fig:vari-eg}. The best variational energy (per site) $-1.2696$ is achieved at $\alpha\approx 0.44$ with the sign convention given in Fig. \ref{fig:peps}(d).

\section{Conclusion}
To conclude, we have systematically investigated a family of resonating AKLT-loop states on square, honeycomb, and kagome lattices. Using a natural PEPS representation, we have shown that the RAL states are critical on square and honeycomb lattices, while on kagome lattice it is a gapped $\mathbb{Z}_2$ spin liquid. We also discussed the realization of the $\mathbb{SO}(3)$ spin-rotation symmetry and clarified its manifestation through explicitly constructing the topological sectors and evaluating the corresponding entanglement spectra on infinitely long cylinders. We considered a one-parameter family of PEPS interpolating between the RAL and RVB states which have distinct symmetry realizations. A critical point has been identified along this interpolation path. Lastly, we have used the RAL states to obtain the best-to-date variational energy for the spin-$1$ Heisenberg model on a kagome lattice.

\section{Acknowledgment}
We thank Ignacio Cirac, Didier Poilblanc, Yi-Zhuang You, Zi Cai, Hong Yao and Fang-Zhou Liu for helpful discussions. WL acknowledges the hospitality of the Max-Planck Institute for Quantum Optics, where part of the work has been performed. HHT gratefully acknowledges Tai-Kai Ng and K. T. Law for hospitality during his visit in Hong Kong University of Science and Technology. This work has been supported by the DFG through SFB-TR12, the EU project SIQS, the DFG Cluster of Excellence NIM, and NSFC 11204149.

\textit{Note added.--} Upon finalizing the manuscript we noticed a recent preprint~\cite{Huang-2013} on closely related topics.

%\setcounter{figure}{0}
%\renewcommand{\thefigure}{A\arabic{figure}}
%\begin{appendix}
%\end{appendix}


\begin{thebibliography}{99}
\bibitem{comment:sym-breaking} One may wonder that by putting polarized spinons on the boundaries we weakly break the $\mathbb{SO}(3)$ symmetry through boundary conditions. However, since the boundaries are infinitely far away, {this} effect on the bulk density matrix is negligibly small. In addition, in case the half-integer spin sector is not a degenerate topological sector, the spinon-pair boundary does not effectively change the parity of entanglement spectrum in the bulk [for example, see ES in the spin-1 RVB region ($\alpha>\alpha_c$) shown in Fig. \ref{fig:ES}].
\end{thebibliography}
\end{document}